\documentclass[vecphys]{svmult}

\usepackage{makeidx}         % allows index generation
\usepackage{graphicx}        % standard LaTeX graphics tool
\usepackage{multicol}        % used for the two-column index
\usepackage[bottom]{footmisc}% places footnotes at page bottom
\makeindex             % used for the subject index
\begin{document}

\title*{Kinematic evolution of the young local associations and the 
Sco-Cen complex}
\author{D. Fern\'andez\inst{1,2}, F. Figueras\inst{1}\and
J. Torra\inst{1}}
\institute{Dpt. d'Astronomia i Meteorologia, IEEC-Univ. de Barcelona, Av. 
Diagonal 647, E-08028 Barcelona, Spain -- \texttt{e-mail: 
david.fernandez@am.ub.es}
\and Observatori Astron\`omic del Montsec, Consorci del Montsec, 
Pla\c ca Major 1, E-25691 \`Ager (Lleida), Spain}
\maketitle
\index{Author1}
\index{Author2}
\index{Author3}
% Use the \index{} command to code your author index

\begin{abstract} 
In this work we propose a scenario for the history of the recent star
formation (during the last 20-30 Myr) in the nearest solar neighbourhood
($\sim$150 pc), from the study of the spatial and kinematic properties of
the members of the so-called young local associations, the Sco-Cen complex
and the Local Bubble, the most important structure observed in the local
interstellar medium (ISM).
\end{abstract}

\section{Introduction}
\label{sec:1}
% Always give a unique label
% and use \ref{<label>} for cross-references
% and \cite{<label>} for bibliographic references
% use \sectionmark{}
% to alter or adjust the section heading in the running head

During the last decade, several young stellar associations have been
discovered within 100 pc of Earth (see \cite{journal1} for a recent
review) thanks to the cross-correlation of the ROSAT and Hipparcos
catalogues, combined with spectroscopy observations. The stars belonging
to these young local associations (herafter, YLA) have ages in the range
from a few to several tens of Myr.

The YLA not only offer insights into the star formation process in
low-density environments, but also have shed light on the substellar
astrophysics, since tens of brown dwarfs have been identified in these
associations. Furthermore, the study of the YLA can also yield important
clues on the recent star formation in our vicinity and its consequences on
the local ISM.

\section{The nearest solar neighbourhood}
\label{sec:2}

\subsection{The Local Bubble}

The local interstellar medium is dominated, in the first 100 pc, by the
so-called Local Bubble (LB). It was discovered at the end of the 1960s as
a soft X-ray diffuse background (SXRB). Rapidly, it was recognized the
existence of an anticorrelation between the X-ray background and the
observed HI column density. The displacement model (see \cite{journal2})
for this structure assumes that the HI irregular local cavity is filled by
an X-ray-emitting plasma, with an emission temperature of $\sim10^6$ K and
a density of $n_e \sim 0.005$ cm$^{-3}$. The contours of the LB were
obtained in \cite{journal3} from NaI absorption measurements towards a
selected set of stellar targets with Hipparcos parallaxes. These
observations allowed the authors to draw maps of the neutral gas
distribution in the local ISM and, in particular, to trace the contours
and extension of the LB with an estimated precision of $\approx\pm$20 pc
in most directions.

Several models have been presented to explain the origin of the LB. In
\cite{contribution1} five conceptions of the LB were reviewed. The maximum
consensus is reach with a scenario where about 10-20 supernovae (SNe)
formed the local cavity and, after that, 1 or 2 SNe reheated the LB a few
Myr later, explaining the present temperature observed for the SXRB (see
\cite{contribution2}).

\subsection{Young local associations}

A decade ago, the number of pre-main sequence stars identified near than
100 pc from the Sun was very low. Nearly all the youngest stars ($<$30
Myr) studied then were located further than 140 pc, in the molecular
clouds of Taurus, Chamaeleon, Lupus, Scorpius-Centaurus and R CrA, all of
them regions of recent star formation. The cross-correlation of Hipparcos
and ROSAT catalogues suddenly changed this view, since a few stars were
identified as very young, but located closer than 100 pc, where there are
not molecular clouds or star forming regions (SFR). Two scenarios were
proposed to explain the existence of these young stars far away from SFR.
In \cite{journal4} it was defended that these stars were born in molecular
clouds, but later they were ejected from them. On the other hand, in
\cite{journal5} it was proposed that these young stars were formed inside
small molecular clouds (or {\it cloudlets}), which were lately dispersed
in the ISM and then they are not detected at present.

During the last decade these nearby, young stars far away from molecular
clouds and SFR have been identified as members of young associations, each
one of them formed by a few tens of members. We have done an intensive
search in the literature, and compiled all the present available data for
these young associations, which are summarized in Table \ref{tab.Assoc}.

\subsection{The Scorpius-Centaurus complex}

The Scorpius-Centaurus (Sco-Cen) complex dominates the 4th galactic
quadrant. It is a region of recent star formation and contains an
important fraction of the most massive stars in the solar vicinity (see
\cite{journal10}). In the 1960s the complex was splitted into three
components: Upper Scorpius (US), Upper Centaurus Lupus (UCL) and Lower
Centaurus Crux (LCC). The most accepted ages for these three associations
were derived in \cite{journal6} from isochrones at the HR diagram:
$\sim$5-6 Myr for US, $\sim$14-15 Myr for UCL and $\sim$11-12 Myr for LCC.
In \cite{journal7} it was suggested that this progression in age was the
result of a sequence of different events of star formation which happened
in the giant molecular cloud which formed the region of Sco-Cen. These
{\it classical} ages have been recently called in question from studies of
the low mass component of the complex. The ages obtained from these
studies are in the range of 8-10 Myr for US and 16-20 Myr for UCL and LCC
(see \cite{journal8}). The three associations show a strong velocity
component in the direction away from the Sun (see Table \ref{tab.Assoc}),
classically associated to the expansion motion of the Gould Belt (see
e.g.\cite{journal9}).

\begin{table}
   \caption{Mean spatial coordinates and heliocentric velocity components
            of the YLA and the Sco-Cen complex. $N$ is the number of known 
            members of each association. Units: $\xi^\prime, \eta^\prime, 
            \zeta^\prime$ in pc; $U, V, W$ in km s$^{-1}$; Age in Myr 
            (1: from \cite{journal6} and 2: from \cite{journal8}).}
   \label{tab.Assoc}
\centering
\begin{tabular}{lrrrrrrrr}
\hline
Association     & $\overline{\xi^\prime}$ &
                  $\overline{\eta^\prime}$ &
                  $\overline{\zeta^\prime}$
                & $\overline{U}$ & $\overline{V}$ & $\overline{W}$
                & Age & $N$ \\
\hline
TW Hya          & $-21_{(22)}$
                & $-53_{(23)}$
                & $21_{(\;\;7)}$
                &  $-9.7_{(4.1)}$ & $-17.1_{(3.1)}$ &  $-4.8_{(3.7)}$
                &  $\sim$8 & 39 \\
Tuc-Hor/GAYA \hspace{-0.4cm}   
                & $-12_{(22)}$
                & $-24_{(11)}$
                & $-34_{(\;\;8)}$
                & $-10.1_{(2.4)}$ & $-20.7_{(2.3)}$ &  $-2.5_{(3.8)}$
                &  20-30 & 52 \\
$\beta$ Pic-Cap & $-9_{(27)}$
                & $-5_{(14)}$
                & $-15_{(10)}$
                & $-10.8_{(3.4)}$ & $-15.9_{(1.2)}$ &  $-9.8_{(2.5)}$
                &  $\sim$12 & 33 \\
$\epsilon$ Cha  & $-47_{(\;\;8)}$
                & $-80_{(14)}$
                & $-25_{(\;\;5)}$
                &  $-8.6_{(3.6)}$ & $-18.6_{(0.8)}$ &  $-9.3_{(1.7)}$
                &  5-15 & 16 \\
$\eta$ Cha      & $-33_{(\;\;2)}$
                & $-80_{(\;\;5)}$
                & $-34_{(\;\;2)}$
                & $-12.2_{(0.0)}$ & $-18.1_{(0.9)}$ & $-10.1_{(0.5)}$
                &  $<$10 & 18 \\
HD 141569       & $-77_{(\;\;3)}$
                & $10_{(\;\;8)}$
                & $64_{(\;\;8)}$
                &  $-5.4_{(1.5)}$ & $-15.6_{(2.6)}$ &  $-4.4_{(0.8)}$
                &  2-8 & 5 \\
Ext. R CrA      & $-97_{(44)}$
                & $-1_{(\;\;4)}$
                & $-30_{(16)}$
                &  $-0.1_{(6.4)}$ & $-14.8_{(1.4)}$ & $-10.1_{(3.3)}$
                &  10-15 & 59 \\
AB Dor          & $6_{(21)}$
                & $4_{(17)}$
                & $-12_{(16)}$
                &  $-7.4_{(3.2)}$ & $-27.4_{(3.2)}$ & $-12.9_{(6.4)}$
                &  75-150 & 40 \\
\hline
US              & $-134\;\;\;\;\;$
                & $-20\;\;\;\;\;$
                & $52\;\;\;\;\;$
                & $-6.7_{(5.9)}$ & $-16.0_{(3.5)}$ & $-8.0_{(2.7)}$
                & 5-6$^1$ & 120 \\
                &  &
                & & & & & 8-10$^2$ & \\
UCL             & $-119\;\;\;\;\;$
                & $-67\;\;\;\;\;$
                & $31\;\;\;\;\;$
                & $-6.8_{(4.6)}$ & $-19.3_{(4.7)}$ & $-5.7_{(2.5)}$
                &  14-15$^1$ & 221 \\
                &  &  &
                & & & & 16-20$^2$ & \\
LCC             & $-62\;\;\;\;\;$
                & $-100\;\;\;\;\;$
                & $10\;\;\;\;\;$
                & $-8.2_{(5.1)}$ & $-18.6_{(7.3)}$ & $-6.4_{(2.6)}$
                &  11-12$^1$ & 180 \\
\hline
\end{tabular}
\end{table}

\section{Kinematic evolution of the young stellar component}

In this section we present the results of our kinematic study of the YLA
and the Sco-Cen complex. This study is based on the integration back in
time of the orbits of these associations, allowing us to study their
origin and their possible influence on the nearest ISM during the last
million years. The integration of the equations of motion has been done
using a realistic galactic potential which includes an axisymmetric
component, the spiral arm perturbation and the central bar contribution
(see \cite{journal11} and \cite{journal13}).

\subsection{Orbits back in time: the origin of the YLA}

\begin{figure}
\centering
\resizebox{7cm}{!}{\includegraphics{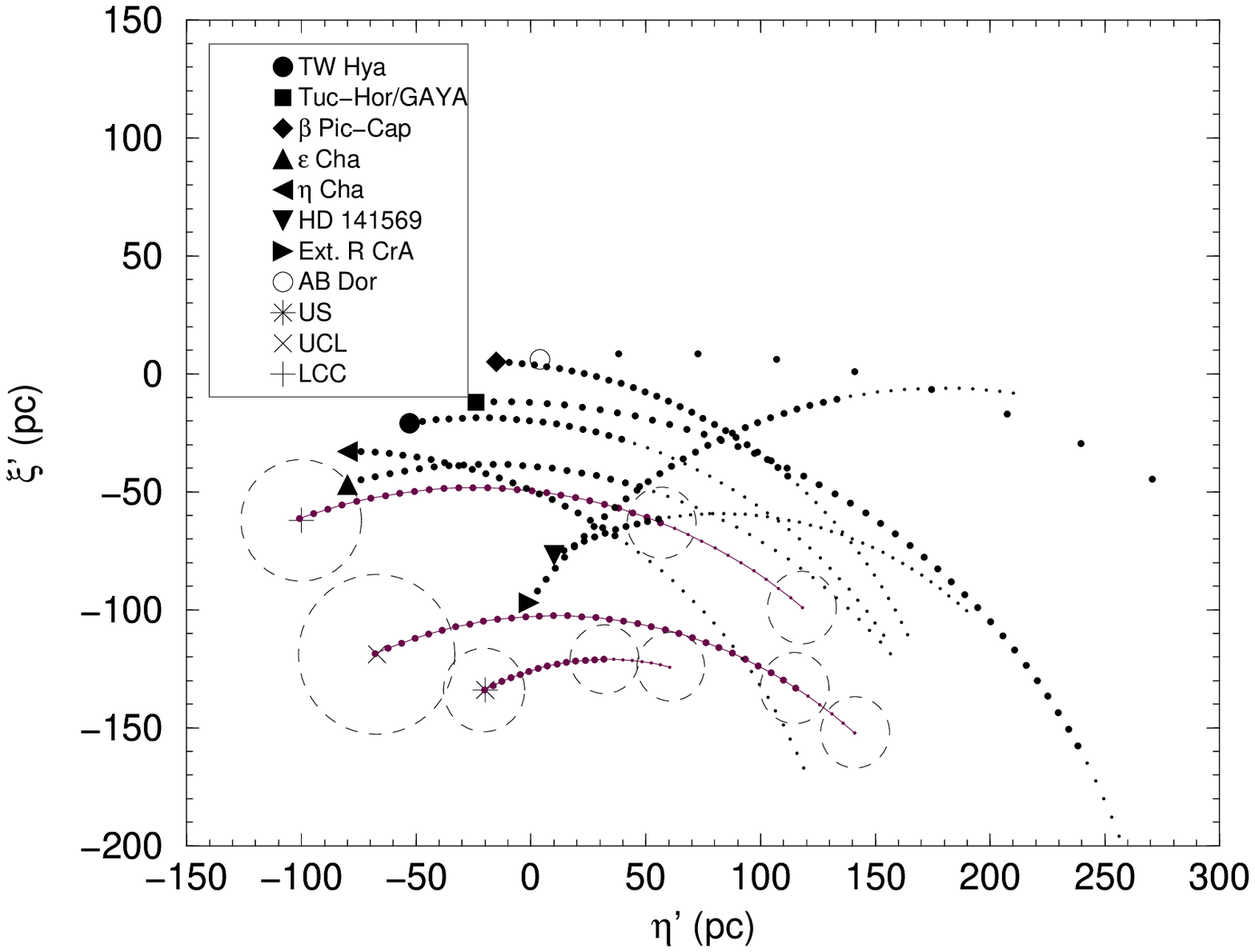}}
\resizebox{9.5cm}{!}{\includegraphics{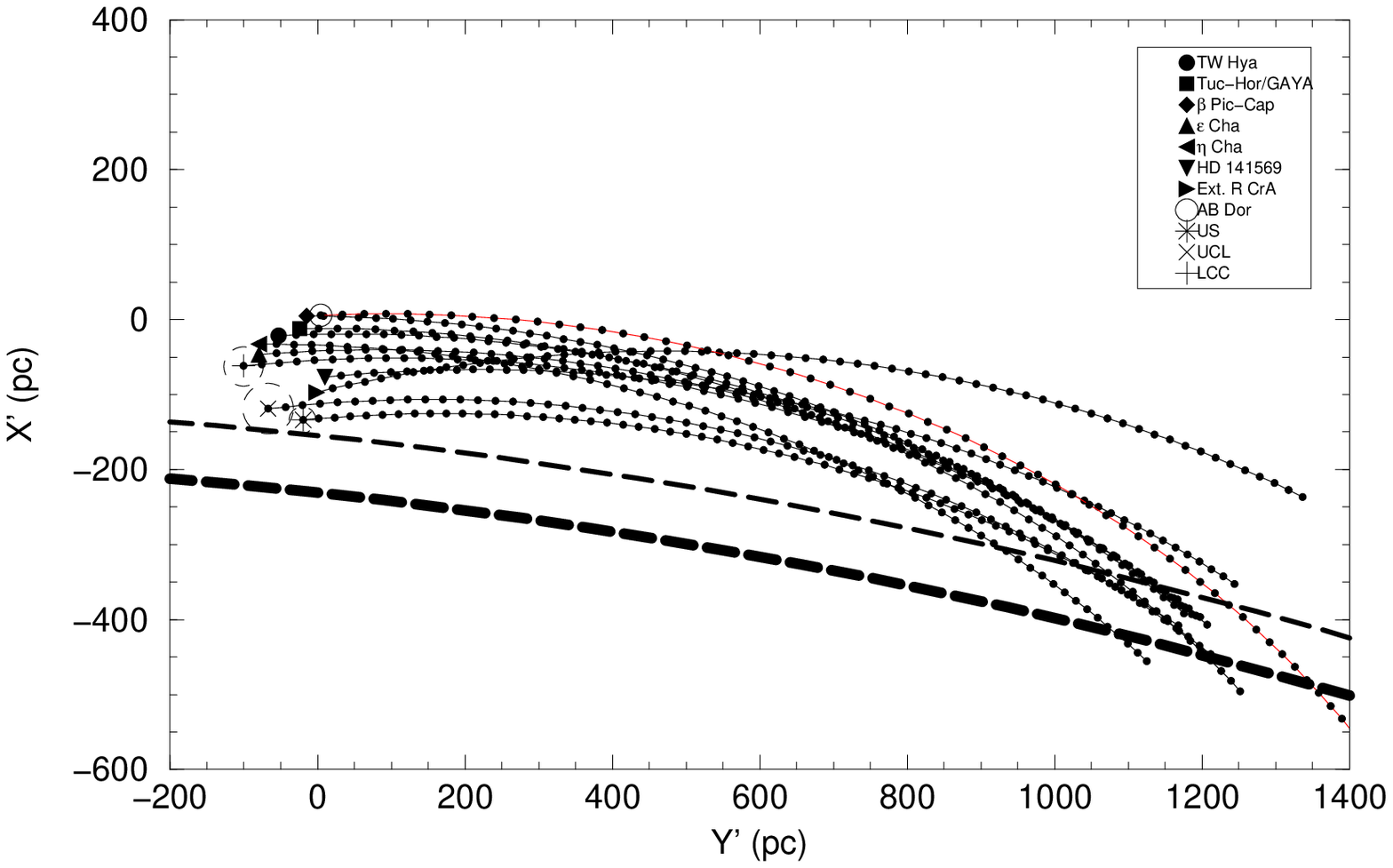}}
\caption{Orbits back in time on the galactic plane to $t = -$30 Myr for 
         the YLA and the Sco-Cen associations, in the LSR reference frame 
         (RF; top) and in the RF which is rotating at the same velocity of 
         the spiral arms (bottom).
         $\xi'$ and $X'$ point to the galactic center, and $\eta'$, $Y'$ 
         point to the direction of the galactic rotation.
         The thick-dashed line shows the position of the minimum of the 
         spiral potential (adopted from \cite{journal12}). In thin-dashed 
         line, the position of the phase of the spiral structure $\psi = 
         10^\mathrm{o}$.}
\label{figure1}
\end{figure}

Studying the orbits back in time for the YLA, we have found as the most
gaudy trend a spatial concentration in the 1st galactic quadrant (see Fig.
\ref{figure1}), in such a way that those stars belonging to the YLA were
concentrated in a region of about 60 x 100 x 40 pc at their birth instant,
whereas at present they are spread inside a region of 120 x 130 x 140 pc
in size (except for AB Dor, which is clearly older than the rest of YLA;
see \cite{journal13}).  Therefore, we can affirm that a star formation
process was triggered in this region of the 1st galactic quadrant, between
5 and 15 Myr ago, producing the YLA. What was the mechanism which
triggered this star formation process? One possibility is the explosion of
one or several close SNe, which should belong to the most important
concentration of O and B stars in our vicinity: the Sco-Cen complex.

It is then of great interest to determine the instant when the minimum
distance between LCC (the closest Sco-Cen association to the YLA during
the last few Myr) and the YLA occurred. For 3 YLA ($\eta$ Cha, $\epsilon $
Cha and Ext. R CrA), minimum distances of $\sim$16, 23 and 73 pc are
obtained for $t \sim -$9 Myr. No clear minima in distance to LCC are found
for the other YLA, but distances of only a few tens of pc are obtained.

\subsection{The YLA and the Local Bubble}

We have found that the trajectories back in time for the YLA have nearly
crossed the center of the Local Bubble (LB) during the last $\sim$5 Myr
(see Fig. 4 in \cite{journal3}). Studying the stellar content of the YLA
(see \cite{journal13} for details), we conclude that it is possible that
one or a few of these associations have sheltered a SN in the recent past
(the last 10 Myr). On the other hand, there is direct evidence for an
explosion of a SN at a distance of $\sim$30 pc from the Sun, $\sim$5 Myr
ago (see \cite{journal14}). Taking into account the trajectories back in
time of the Sco-Cen associations that we have computed, it is not probable
that this SN could exploded in one of them, since not even LCC has
approached as much to the solar neighbourhhod in the past. Then, several
pieces of the same puzzle seem to support the existence of a SN explosion
in the nearest solar neighbourhhod ($\sim$30 pc) $\sim$5 Myr ago, from a
parent star belonging to one of the YLA, probably Tuc-Hor/GAYA or Ext. R
CrA. This close and recent SN would have been responsible for the
reheating of the gas inside the LB needed to achieve the observed
temperatures in the soft X-ray diffuse background at present.

\section{A scenario for the recent star formation in our vicinity}

To complete the scenario outlined in the previous section, we should study
the origin of the Sco-Cen associations. Several scenarios have been
proposed to explain the origin of this region (see \cite{journal8}), but a
careful analysis (see \cite{journal13}) allows to conclude that the most
promising is the impact with a spiral arm. This scenario is supported by
several theoretical and observational facts (see \cite{journal13} and
\cite{thesis1}). Taking this as an hypothesis, the history of the star
formation in the nearest solar neighbourhood during the last few tens of
million years would have been written as follows. 30 Myr ago, the giant
molecular cloud parent of Sco-Cen was placed in a position of the galactic
plane with coordinates $(X^{\prime},Y^{\prime}) \sim (-400,1200)$ pc. The
arriving of the potential minimum of the inner spiral arm triggered the
star formation in the region, disturbing at the same time the cloud
motion, which began to move with a velocity vector directed to the
galactic antirotation and away from the galactic center (see Fig.  
\ref{figure1}; just the expected motion after an interaction with the
spiral arm for a position outside the corrotation radius). The compression
due to the spiral arm did not necessarily trigger the star formation in
the whole cloud, but perhaps only in those regions with the largest
densities. This would be favoured by the smaller relative velocity between
the shock wave and the RSR (12-13 km s$^{-1}$, see \cite{journal13}). The
regions where the star formation began must be those which generate UCL,
LCC and, probably, the Tuc-Hor/GAYA association, which were born nearly at
the same time, about 16-20 Myr ago. The stellar winds of the first massive
stars began to compress the gas of the neighbouring regions, maybe
producing their fragmentation in small molecular clouds which moved away
from the central region of the parent molecular cloud. 9 Myr ago, a SN in
LCC could trigger the star formation in these small molecular clouds,
giving birth the majority of the YLA, as we have seen in the previous
section. The stellar winds of the just born stars rapidly expeled the
remaining gas from these small clouds (the {\it cloudlets} proposed by
\cite{journal5}), completely erasing every trace of them and promoting
that we do not observe gas in these regions at present time.  Later, as it
was proposed in \cite{journal15}, the shock front of a SN in UCL would
have triggered the star formation in US about 6 Myr ago. Only 1.5 Myr ago,
the most massive star in US would have exploded as SN and its shock front
would be reaching now the molecular cloud of $\rho$ Oph, triggering there
the beginning of the star formation process. This would complete our
scenario for the recent star formation in the nearest solar neighbourhood.

\vspace{0.2cm}

\noindent {\it Acknowledgements.} This work has been supported by the
CICYT under contracts AYA2003-07736 and AYA2006-15623-C02-02.

%
%
% BibTeX users please use
% \bibliographystyle{}
% \bibliography{}
%
% Non-BibTeX users please follow the syntax
% the syntax of "referenc.tex" for your own citations
%%%%%%%%%%%%%%%%%%%%%%%% referenc.tex %%%%%%%%%%%%%%%%%%%%%%%%%%%%%%
% sample references
% "physics"
%
% Use this file as a template for your own input.
%
%%%%%%%%%%%%%%%%%%%%%%%% Springer-Verlag %%%%%%%%%%%%%%%%%%%%%%%%%%

%
% BibTeX users please use
% \bibliographystyle{}
% \bibliography{}

\begin{thebibliography}{99.}
% Use the following syntax and markup for your references
%
% Monographs

%\bibitem{monograph} H. Ibach, H. L\"uth: \textit{Solid-State
%Physics}, 2nd edn (Springer, Berlin Heidelberg New York 1996) pp 45--56

% Contributed Works

%\bibitem{contribution} D.M. MacKay: Visual stability and voluntary eye
%movements. In: \textit{Handbook of Sensory Physiology}, vol 3, ed by R.
%Jung, D.M. MacKay (Springer, Berlin Heidelberg New York 1973) pp
%307--331

% Journal

\bibitem{journal11} R. Asiain, F. Figueras, J. Torra: A\&A \textbf{350},
434 (1999)

\bibitem{journal7} A. Blaauw: ARA\&A \textbf{2}, 213 (1964)

\bibitem{contribution2} D. Breitschwerdt, D.P. Cox: Is the Local Bubble 
Dead?. In: \textit{How does the Galaxy work? A Galactic Tertulia with Don 
Cox and Ron Reynolds}, ed by E.J. Alfaro, E. P\'erez, J. Franco 
(A\&SS Library, vol. 315, 2004), 391--401

\bibitem{contribution1} D.P. Cox: Modeling the Local Bubble. In: 
\textit{Proceedings of the IAU Colloquium 166: The Local Bubble and 
Beyond}, ed by D. Breitschwerdt, M.J. Freyberg, J. Tr\"umper (Lecture 
Notes in Physics, vol. 506, 1998), 121--126

\bibitem{journal5} E.D. Feigelson: ApJ \textbf{468}, 306 (1996)

\bibitem{journal12} D. Fern\'andez, F. Figueras, J. Torra: A\&A 
\textbf{372}, 833 (2001)

\bibitem{journal13} D. Fern\'andez, F. Figueras, J. Torra: A\&A, submitted 
(2007)

\bibitem{thesis1} D. Fern\'andez: Estructura espacial y cinem\'atica de la
componente estelar joven en el entorno solar. PhD Thesis, Universitat de
Barcelona, Barcelona (2005)

\bibitem{journal6} E. de Geus, P.T. de Zeeuw, J. Lub: A\&A \textbf{216}, 
44 (1989)

\bibitem{journal14} K. Knie, G. Korschinek, T. Faestermann et al.: Phys. 
Rev. Lett. \textbf{81}, 18 (1999)

\bibitem{journal3} R. Lallement, B.Y. Welsh, J.L. Vergely, F. Crifo, D.M. 
Sfeir: A\&A \textbf{411}, 447 (2003)

\bibitem{journal15} T. Preibisch, H. Zinnecker: AJ \textbf{117}, 2381 
(1999)

\bibitem{journal8} M.J. Sartori, J.R.D. L\'epine, W.S. Dias: A\&A 
\textbf{404}, 913 (2003)

\bibitem{journal2} S.L. Snowden, D.P. Cox, D. McCammon, W.T. Sanders: ApJ 
\textbf{354}, 211 (1990)

\bibitem{journal4} M. Sterzik, R. Durisen: A\&A \textbf{304}, L9 (1995)

\bibitem{journal9} J. Torra, D. Fern\'andez, F. Figueras: A\&A
\textbf{359}, 82 (2000)

\bibitem{journal10} P.T. de Zeeuw, R. Hoogerwerf, J.H.J. Bruijne et al.:
AJ \textbf{117}, 354 (1999)

\bibitem{journal1} B. Zuckerman, I. Song: ARA\&A \textbf{42}, 685 (2004)

% Theses

\end{thebibliography}

%%%%%%%%%%%%%%%%%%%%%%%%%%%%%%%%%%%%%%%%%%%%%%%%%%%%%%%%%%%%%%%%%%%%%%  }

%%%%%%%%%%%%%%%%%%%%%%%%%%%%%%%%%%%%%%%%%%%%%%%%%%%%%%%%%%%%%%%%%%%%%%

\printindex
\end{document}